\documentclass[letter]{aa}
\usepackage{txfonts}
\usepackage{graphicx}

\begin{document}

\title{Evidence for bipolar jets from the optical spectra of the prototypical
symbiotic star Z~Andromedae \thanks{Based on observations collected at 
Tartu Observatory, Estonia}}

\author{M. Burmeister\inst{1,2}
  \and L. Leedj\"{a}rv\inst{1}}

\offprints{M. Burmeister, \email{ mari@aai.ee}}

\institute{Tartu Observatory, 61602 T\~{o}ravere, Estonia
  \and Institute of Theoretical Physics, University of Tartu,
  T\"{a}he 4, 51010 Tartu, Estonia} 

\date{Received 24 October 2006 / Accepted 3 November 2006}

\abstract{}
{We have studied optical spectra of the symbiotic star Z~And,
obtained during its latest outburst started in April 2006, with the aim of
finding changes in the spectrum yielding clues to the nature of the hot
component and its outbursts.} {The spectroscopic observations of Z~And
have been made
using the 1.5-meter telescope at the Tartu Observatory, Estonia, and
processed in a standard way.} {We have found high velocity satellites
to the hydrogen Balmer emission lines. Starting from July 30, 2006, weak
additional emission components at velocities of about $\pm$ 1150\,km\,s$^{-1}$ 
were detected. Their appearance near the outburst maximum and similarity to
the emission features in another symbiotic star Hen~3-1341 imply fast 
collimated outflows from the hot component of Z~And. This finding is 
consistent with the earlier results by several authors that symbiotic stars 
can emit bipolar jets at certain stages of their outbursts. A significant 
decrease in the temperature of the hot component in initial stages of the 
outburst was detected by the disappearance of the high excitation emission lines 
from the spectrum.}
{}


\titlerunning{Bipolar jets from Z And}
\maketitle

\section{Introduction}

Z~Andromedae is considered to be the prototype for the class of symbiotic
stars
(Kenyon \cite{ken86}; Corradi et al. \cite{corr03}) -- interacting binary 
stars consisting of a red giant
and a hot compact companion, mostly a white dwarf. The hot component ionises
part of the wind of the red giant, thus giving rise to the characteristic
"combination" spectrum where high excitation emission lines are superimposed
on the cool giant's spectrum. 
\object{Z~And} has an orbital period of 757.5 days (Miko{\l}ajewska \& Kenyon
\cite{miken96}) and an orbital inclination of $47\degr \pm 12\degr$ (Schmid
\& Schild \cite{schsch}). Its light curve represents classical symbiotic
star outbursts, demonstrating brightening of the star by 2--3 magnitudes in
the visual region in every few years. Z~And is also the only symbiotic star
for which coherent optical oscillations with the period of 28 minutes have
been detected, most likely indicating the presence of a strongly magnetic WD
(Sokoloski \& Bildsten \cite{sokbi}). Finally, Z~And is among about 10 known
symbiotic stars, producing collimated outflows or jets (Brocksopp et
al. \cite{bro04}; Leedj\"arv \cite{leed04}). Extended collimated outflows
perpendicular to the orbital plane were detected from radio images during
the 2000--2002 outburst (Brocksopp et al. \cite{bro04}).

Behaviour of Z~And during the outburst and the following quiescence
in 2000--2003 was extensively studied and analysed by Sokoloski et al.
(\cite{sok06}). They proposed to call Z~And and possibly other classical
symbiotic stars 'combination novae', indicating that both dwarf nova-like
accretion disk instability and nova-like nuclear shell burning are involved
in the outbursts of classical symbiotic stars. According to the visual
photometric observations from the AAVSO database, a new outburst of Z~And 
began in early June 2004. Having hardly recovered from this outburst,
the star started a new activity cycle in April 2006.
In the present paper we describe and analyse the
optical spectra of Z~And, obtained mostly during decline from the optical
maximum in 2006. Section 2 describes the observations and in Sect. 3 we
present and analyse evidence for fast bipolar outflows during the outburst
of Z~And. Section 4 contains conclusions.

\section{Observations}

Spectroscopic observations of Z~And have been carried out at the Tartu
Observatory, Estonia, using the 1.5-meter telescope equipped with
the Cassegrain grating spectrograph. Two different CCD cameras from
the SpectraSource Instruments were used until March 2006. Later on
the spectra are registered with the Andor Technologies CCD camera 
Newton USB-207 with $400\times1600$ chip, pixel size
$16\times16\,\mu$m, Peltier cooled. A few occasional spectra of Z~And 
have been obtained since September 1997, mostly in the region of H$\alpha$.
A more systematic monitoring started on July 30, 2006, after
detecting the signs for bipolar jets.

Most of the red spectra cover the wavelength region 6480$-$6730\,\AA,
with linear dispersion about $0.2\,\AA\,\mathrm{pixel^{-1}}$. Blue
spectra are mostly taken in the 4600$-$5060\,\AA~ region with 
$\sim 0.3\,\AA\,\mathrm{pixel^{-1}}$. Some of the lower dispersion spectra
(about $0.5\,\AA\,\mathrm{pixel^{-1}}$ and $0.6\,\AA\,\mathrm{pixel^{-1}}$,
respectively) are also used.
The spectra were reduced using the software package MIDAS provided by ESO. After standard procedures of subtracting the bias and sky background, the
spectra were reduced to the wavelength scale using the Th-Ar hollow cathode
lamp spectrum for calibration. Thereafter, the spectra were normalised to
the continuum, and positions, peak intensities, and equivalent widths of the
emission lines were measured.

\section{Results and discussion}

\begin{figure}
\centering
\resizebox{\hsize}{!}{\includegraphics[angle=270]{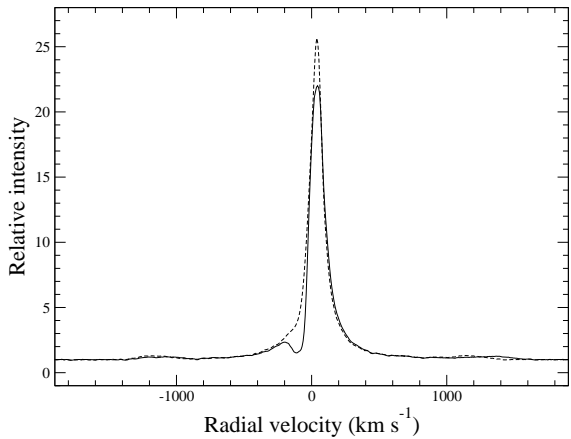}}
\caption {H$\alpha$ line profiles from August 7 (solid line)
and September 10 (dotted line).
The former shows an absorption component.}
\label{fig:ha2}
\end{figure}

Both H$\alpha$ and H$\beta$ lines have a central emission component
at radial velocities of about 30~km\,s$^{-1}$.
The shape of the profiles is asymmetric with the blue wing distorted
by absorption. On August 7 and 9, the blue step of the
H$\alpha$ line deepens into an absorption component (Fig.~\ref{fig:ha2}).
A similar absorption is seen in H$\beta$ on September 1.

\begin{figure}
\centering
\resizebox{\hsize}{!}{\includegraphics{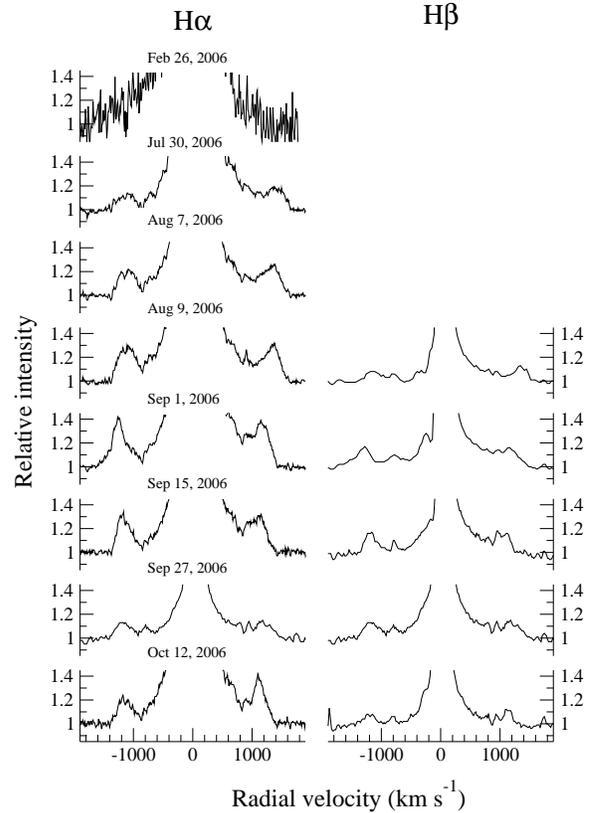}}
\caption {Time evolution of the jet components around the H$\alpha$
and H$\beta$ lines. Our last spectrum before the outburst
from February 26, although more noisy, does not show additional emission
components.}
\label{fig:hahb}
\end{figure}

The most striking features of these Balmer lines, however, are
the satellite emission components (see Fig.~\ref{fig:hahb}),
detected for the first time on July 30, 2006 in the H$\alpha$ line.
We are not aware of observations of
similar features in the spectrum of Z~And before. On September 14,
2006, a telegram on the detection of jets in Z~And was published by Skopal \&
Pribulla (\cite{skopri}).
Very similar additional emission features were discovered
in the spectra of the symbiotic stars \object{Hen 3-1341}
by Tomov et al. (\cite{tomov}), and \object{StH$\alpha$~190}
by Munari et al. (\cite{mun01}), who interpreted them to be emitted by 
the collimated bipolar jets.

\begin{table}
\caption{Radial velocities of the jet components of H$\alpha$
and H$\beta$ line in km\,s$^{-1}$}
\label{table:hv}
\centering
\begin{tabular}{llllll}                                  \hline\hline
Date &  JD -  &
  \multicolumn{2}{c}{H$\alpha$}  &
  \multicolumn{2}{c}{H$\beta$}      \\
2006  &  2\,400\,000 & Blue & Red & Blue & Red \\  \hline
  Jul 30  &  53947.5  &  -1114  &  1391  &         &        \\
  Aug 7   &  53955.5  &  -1115  &  1271  &         &        \\
  Aug 9   &  53957.4  &  -1106  &  1275  & -969    &  1376  \\
  Aug 10  &  53958.5  &  -1134  &  1280  &         &        \\
  Sep 1   &  53980.4  &  -1243  &  1165  &  -1288  &  1160  \\
  Sep 10  &  53989.4  &  -1150  &  1157  &  -1180  &  1084  \\
  Sep 12  &  53991.4  &  -1193  &  1145  &  -1217  &  1127  \\
  Sep 13  &  53992.5  &  -1180  &  1067  &  -1191  &  1061  \\
  Sep 15  &  53994.4  &  -1156  &  1059  &  -1176  &  1039  \\
  Sep 21  &  54000.4  &  -1081  &  1100  &  -1070  &  1098  \\
  Sep 24  &  54003.5  &  -1114  &  1087  &  -1162  &  1137  \\
  Sep 27  &  54006.4  &  -1115  &  1107  &  -1150  &  1134  \\
  Oct 6   &  54015.3  &  -1106  &  1110  &         &        \\
  Oct 12  &  54021.4  &  -1124  &  1108  &  -1196  &  1134  \\  \hline
  Mean    & & -1138$\pm$41 & 1166$\pm$96 & -1160$\pm$82 & 1135$\pm$88 \\ \hline
\end{tabular}
\end{table}

We fitted these emission components
with a Gaussian to find their radial velocities. The results
are given in Table~\ref{table:hv}.
On average the RVs are about $\pm 1150\,\mathrm{km\,s^{-1}}$.
Given the asymmetric shape of the additional components, the accuracy of
those velocities is no better than about 40$-$50 km\,s$^{-1}$.
If we assume that the jets are perpendicular to the orbital plane
and the inclination of the orbit is $47\degr$, the true velocities
of the jets are close to $\pm 1700\,\mathrm{km\,s^{-1}}$.
\begin{figure}
\resizebox{\hsize}{!}{\includegraphics[angle=270]{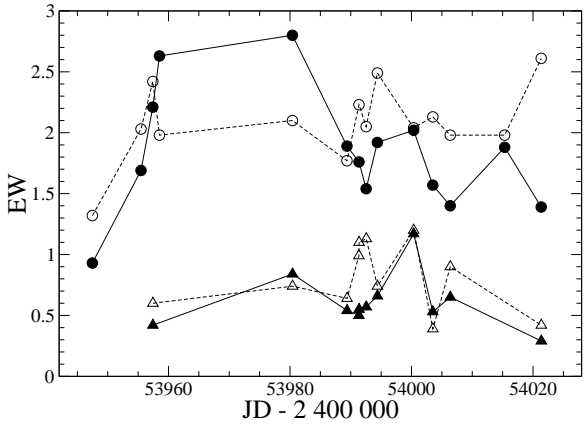}}
\caption {EWs of the blue (filled circles) and red (empty circles)
jet component of the H$\alpha$ line and of the blue (filled triangles)
and red (empty triangles) jet component of the H$\beta$ line.}
\label{fig:ew}
\end{figure}

The equivalent widths of the supposed jet components form
about 2\% of that of the central
emission, but these estimates are uncertain, as
the small emissions are partly blended in the central component.
The measured values of EWs are plotted in Fig.~\ref{fig:ew}. 

At least in some of the red spectra, weak [\ion{N}{II}] 6548 and 6584~$\AA$
emission lines can be distinguished at the low-velocity sides of
H$\alpha$ jet components.
Where measurable, the velocities of the [\ion{N}{II}] lines occur at
about $-$45~km\,s$^{-1}$.

According to the AAVSO light curve, the optical maximum of Z~And took place
in early July 2006 (around JD 2\,453\,920 $-$ 930). Our last optical
spectrum before the outburst was made on February 26, 2006. No signs of jets
are seen in this spectrum, so the jets must have been formed
at some time between February and July.
For
most of the time of our observations, the jet features in the spectrum have
been rather stable. Smaller EWs on July 30 possibly indicate the early phase
of jet development. 

Such an appearance of the jets at the time of the optical maximum fits
the behaviour of another symbiotic star \object{Hen~3-1341} as described 
by Munari et al. (\cite{mun05}) well. In general, collimated bipolar
jets are observed from
several types of stars, such as, e.g., X-ray binaries, supersoft X-ray sources,
young stellar objects, etc., and of course, from active galactic nuclei. The
standard scenario for the production of jets (e.g., Livio \cite{livio};
Lynden-Bell \cite{lynd}, and
many later references) includes an accretion disc that is threaded by a
vertical magnetic field. In addition, an energy or wind source associated
with the central accreting object is needed. This requirement explains, 
for instance, the absence of
jets from magnetic cataclysmic variables, in spite of the presence of
an accreting white dwarf and magnetic field. The behaviour of Hen~3-1341 well
demonstrates that the jets are emitted only in the early and the brightest
phase of the outburst, and that they are fed by the wind from the
outbursting component (Munari et al. \cite{mun05}). The outburst of
Hen~3-1341, however, lasted longer (from 1998 to 2004) than those of Z~And
usually do.

\begin{figure}
\centering
\resizebox{\hsize}{!}{\includegraphics{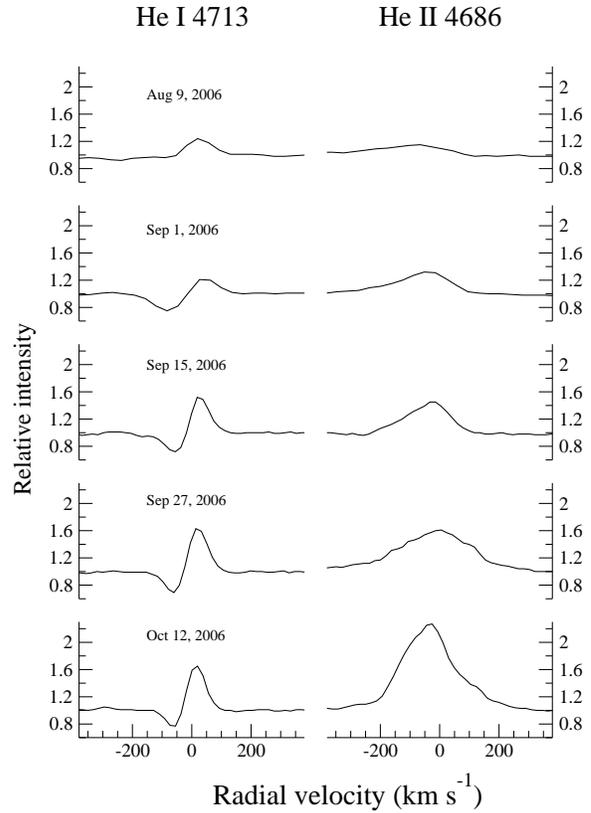}}
\caption {Time evolution of the \ion{He}{I} 4713 and \ion{He}{II} 4686 lines,
representing matter outflow and outburst state, respectively.
The profiles are given on the same scale for comparison.}
\label{fig:he1he2}
\end{figure}

\begin{figure}
\resizebox{\hsize}{!}{\includegraphics[angle=270]{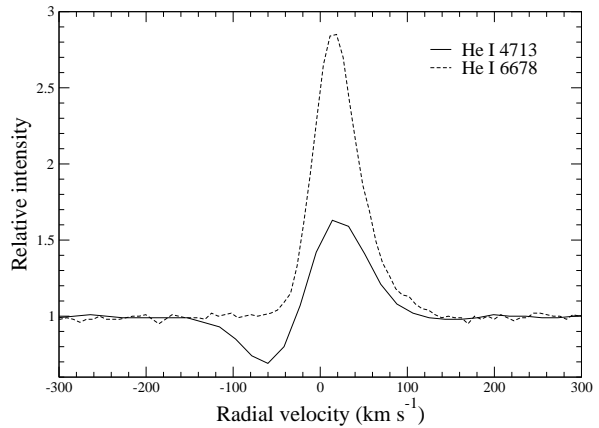}}
\caption {Example of an \ion{He}{I} 4713 line and \ion{He}{I} 6678 line
from September 27, 2006.}
\label{fig:4713_6678}
\end{figure}

The wings of the H$\alpha$ main emission component extend to about $\pm
1000$ km\,s$^{-1}$, which may indicate fast wind from the outbursting hot
component. On the other hand, indications of low-velocity outflow can be seen in
the lines of \ion{He}{I} 4471, 4713, and 5016, which present P Cygni profiles
with deep absorption components. The development of the P~Cygni profile 
of \ion{He}{I} 4713 is seen in Fig.~\ref{fig:he1he2}.
Velocity of the outflow measured from these lines is
between $-50$ and $-100$~\,km\,s$^{-1}$.
This is comparable to the velocity $-$90
kms$^{-1}$ found by Skopal et al. (\cite{skopal}) during the 2000--2003
outburst. At the same time, the \ion{He}{I} 4922 and 6678 lines are fully in 
emission or have only a very weak absorption component.
Figure~\ref{fig:4713_6678} 
presents an example.

\begin{figure}
\resizebox{\hsize}{!}{\includegraphics[angle=270]{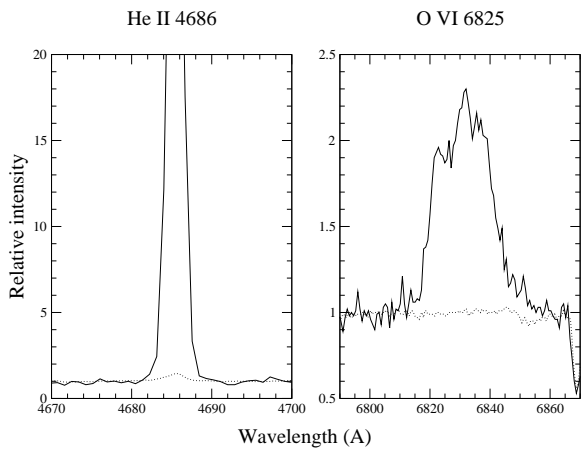}}
\caption {\emph{Left:} \ion{He}{II} 4686 line of Z~And from September 27, 1997
(solid line) and September 27, 2006 (dotted line). \emph{Right:} Raman
scattered line O~VI 6825 from September 2, 1997 (solid line) and
September 27, 2006 (dotted line). Disappearance of the high ionization
emission lines refers to an outburst state.}
\label{fig:HejaO}
\end{figure}

The present outburst of Z~And is similar to that of Hen~3-1341 also by the
behaviour of other emission lines. It is known that high excitation emission
lines tend to weaken during the outbursts of symbiotic stars. In Hen~3-1341,
the He II 4686 and the Raman scattered O VI line at 6825
disappeared completely. Similar changes in the spectrum of Z~And during the
2000--2003 outburst were reported by Skopal et al. (\cite{skopal}) and
Sokoloski et al. (\cite{sok06}). We also an detected almost complete 
disappearance of the Raman scattered line at 6825 and of the 
He II 4686 line in the spectrum of Z~And during the present 
outburst in 2006.
Figure~\ref{fig:HejaO} presents the comparison of those spectral lines from
the quiescence in September 1997 and from the outburst in September 2006.
In Fig.~\ref{fig:he1he2}, the gradual reappearance of the \ion{He}{II} 4686 line
is seen while the brightness of the star decreases.

\begin{table}
\caption{Effective temperatures of the hot star estimated by the Iijima method}
\label{table:temp}
\centering
\begin{tabular}{llr}                                  \hline \hline
  Date  & JD -          &  Temperature       \\
  2006  &  2\,400\,000  &
       (K)             \\  \hline
  Aug 9   &  53957.438    &  76\,000          \\
  Sep 10  &  53989.398    &  91\,000          \\
  Sep 12  &  53991.346    &  80\,000          \\
  Sep 13  &  53992.549    &  75\,000          \\
  Sep 15  &  53994.394    &  81\,000          \\
  Sep 21  &  54000.363    &  87\,000          \\
  Sep 24  &  54003.519    &  91\,000          \\
  Sep 27  &  54006.388    &  100\,000         \\
  Oct 12  &  54021.368    &  114\,000         \\  \hline
\end{tabular}
\end{table}

We made an attempt to estimate the effective temperature of the hot
component during our observations by using the
method proposed by Iijima (\cite{iijima}). Like Sokoloski et al.
(\cite{sok06}) we used equivalent widths of
H$\beta$ and \ion{He}{II} 4686 instead of fluxes and neglected the
\ion{He}{I} 4471 line. The temperatures in August and September tend
to be low, falling in the range 75\,000--90\,000~K, and
begin to rise at the end of September. The results are given
in Table~\ref{table:temp}. A similar decline of $T_{\mathrm{hot}}$ at the 
brightness maximum and its gradual increase after the maximum was 
found by Sokoloski et al. (\cite{sok06}) during the 2000--2003 outburst. 
On the contrary, $T_{\mathrm{hot}}$ increased from the quiescence value of
$\la 150\,000$ K to about 180\,000 K during the smaller scale short outburst
in 1997. Those facts imply similarity with another classical symbiotic star
\object{AG~Dra} in which Gonz{\'a}lez-Riestra et al. (\cite{gonz99}) have
distinguished between cool and hot outbursts, based on the changes in the
\ion{He}{II} Zanstra temperature. The outbursts of Z~And in 2000
and 2006 can be considered as cool ones and that in 1997 as hot outburst.
It might be of interest to note that AG~Dra has also been in outburst
since July 2006. We will describe spectroscopic behaviour of AG~Dra in
another forthcoming paper. Here we only note that during the present
outburst the \ion{He}{II} 4686 line and the Raman scattered \ion{O}{VI} line
6825 also became very weak in the spectrum of AG~Dra. This is different from
smaller scale outbursts of AG~Dra in the late 1990s -- early 2000s when all 
the emission lines became stronger (e.g., Tomov \& Tomova \cite{tom02}; 
Leedj{\"a}rv et al. \cite{leedb04}).

\section{Conclusions}

From our observations of the prototypical symbiotic star Z~And we can
conclude the following:

(1) The additional emission components of the hydrogen H$\alpha$ and
H$\beta$ lines at $\pm 1150$ km\,s$^{-1}$ indicate ejection of bipolar jets
starting from late July 2006, and persisting for at about three
months, at least. Taking into account the orbital inclination 47$\degr$, ejection
velocity of the jets would be about 1700 km\,s$^{-1}$.

(2) Although Z~And has been extensively studied over the past decades, no
such jets were detected from the optical spectra before.
We have some spectra in the H$\alpha$ region near the
brightness maximum of the 2004 outburst,
but no additional emission components can be found from these.
According to Skopal et al. (\cite{skopal}), no signs of jets could be seen
during the maximum of the 2000--2003 outburst.
Spectra by Sokoloski et al.\ (\cite{sok06}) from the same time
have too low resolution to confirm the existence or absence of
additional emissions.
We also have one spectrum from August 25, 2001, which was taken almost
at the same time when Brocksopp et al. (\cite{bro04}) discovered
the radio jets from Z~And. In our spectrum, the jets are not seen.
This most likely
means that the jets are transient phenomenon, emitted only for a short time
in the brightest phase of the outbursts, thus confirming the standard 
scenario for jet ejection in which an additional source of energy or wind 
is needed besides the accreting central body and magnetic field.

(3) The high excitation lines \ion{He}{II} 4686 and Raman scattered
\ion{O}{VI} 6825 almost disappeared from the spectrum of Z~And in the
early phase of the outburst, confirming that the temperature of the hot
component decreased to about 75\,000 K, as was found by the Iijima method. In
September 2006, about 3--4 months after the start of the outburst, the
\ion{He}{II} 4686 line started to become stronger again.

The ongoing outburst of Z~And has shown that new aspects can be
found in the behaviour of such well studied symbiotic stars. Together with the
combination nova model proposed by Sokoloski et al. (\cite{sok06}), new
clues for understanding the outbursts of AG~Dra, Hen~3-1341, and other
classical symbiotic stars can be obtained. Continuing monitoring of the
ongoing outburst of Z~And is strongly encouraged.

\begin{acknowledgements}
The authors thank Dr. Kalju Annuk for taking some of the spectra of Z~And
and for useful comments on the manuscript of the paper.
We also thank our referee, Dr. Michael Bode, for his suggestions.
The present study was supported by the Estonian Ministry of Education and
Research under the target financed project 0062464S03 "Structure, chemical
composition, and evolution of stars" and by the Estonian Science Foundation
grant No 6810.
\end{acknowledgements}

\end{document}